%% file: main.tex
\documentclass[acmsmall,screen]{acmart}

\setcopyright{none}
\copyrightyear{2025}
\acmYear{2025}
\acmISBN{}
\acmDOI{}

\input{packages.tex}

\input{defs.tex}

\title{Typed Embedding of \mk for Functional Conversion}

\author{Igor Engel}
\email{igorengel@mail.ru}
\affiliation{%
  \institution{JetBrains Research}
  \city{Munich}
  \country{Germany}
}
\affiliation{%
  \institution{Constructor University}
  \city{Bremen} 
  \country{Germany}
}

\author{Ekaterina Verbitskaia}
\email{ekaterina.verbitskaya@jetbrains.com}
\affiliation{%
  \institution{JetBrains Research}
  \city{Amsterdam}
  \country{Netherlands}
}
\affiliation{%
  \institution{Constructor University}
  \city{Bremen}
  \country{Germany}
}

\renewcommand{\shortauthors}{Engel et al.}

\input{abstract.tex}



\begin{document}

\settopmatter{printacmref=false}
\settopmatter{printfolios=true}
\renewcommand\footnotetextcopyrightpermission[1]{}
\pagestyle{fancy}
\fancyfoot{}
\fancyfoot[R]{miniKanren'25}
\fancypagestyle{firstfancy}{
  \fancyhead{}
  \fancyhead[R]{miniKanren'25}
  \fancyfoot{}
}
\makeatletter
\let\@authorsaddresses\@empty
\makeatother

\maketitle

\thispagestyle{firstfancy}

\input{introduction.tex}
\input{background.tex}

\input{embedding.tex}

\input{conversion.tex}

\input{experiments.tex}
\input{conclusion.tex}

\bibliographystyle{plain}
\bibliography{main}

\end{document}

%% file: packages.tex
\usepackage{xspace}

\usepackage[final]{listings}
\usepackage{libertine}

%% file: defs.tex
\newcommand{\mk}{\textsc{miniKanren}\xspace}
\newcommand{\micro}{\textsc{microKanren}\xspace}
\newcommand{\merc}{\textsc{Mercury}\xspace}

\newcommand{\haskell}{\textsc{Haskell}\xspace}
\newcommand{\ocaml}{\textsc{OCaml}\xspace}
\newcommand{\ocanren}{\textsc{OCanren}\xspace}

\definecolor{lstKeyword}{HTML}{000080}
\definecolor{lstBG}{HTML}{FFFFFF}
\definecolor{lstString}{HTML}{008000}
\definecolor{lstNumber}{HTML}{0000FF}
\definecolor{lstComment}{HTML}{808080}
\definecolor{lstLineNumber}{HTML}{999999}
\lstdefinestyle{customhask}{ 
    belowcaptionskip=1\baselineskip, 
    breaklines=true, 
    xleftmargin=\parindent, 
    language=haskell, 
    backgroundcolor=\color{lstBG},
    commentstyle=\itshape\color{lstComment},
    keywordstyle=\bfseries\color{lstKeyword},
    stringstyle=\color{lstString},
    basicstyle=\scriptsize\ttfamily,
    breakatwhitespace=false,        
    breaklines=false,               
    captionpos=b,                    
    keepspaces=true,              
    showspaces=false,
    showstringspaces=false,
    showtabs=false, 
    tabsize=4,
    columns=flexible
} 
\lstdefinestyle{normalhask}{ 
    belowcaptionskip=1\baselineskip, 
    breaklines=true, 
    xleftmargin=\parindent, 
    language=haskell, 
    backgroundcolor=\color{lstBG},
    commentstyle=\itshape\color{lstComment},
    keywordstyle=\bfseries\color{lstKeyword},
    stringstyle=\color{lstString},
    basicstyle=\normalsize\ttfamily,
    breakatwhitespace=false,        
    breaklines=false,               
    captionpos=b,                    
    keepspaces=true,              
    showspaces=false,
    showstringspaces=false,
    showtabs=false, 
    tabsize=4,
    columns=flexible
} 
\usepackage{xparse} 
\usepackage{xspace}
 
\newcommand{\includecode}[2][hask]{\lstinputlisting[escapechar=@, style=custom#1,xleftmargin=6.0ex]{#2}} 
\newcommand{\includelines}[4][hask]{\lstinputlisting[escapechar=, style=custom#1, firstline=#3, lastline=#4,xleftmargin=6.0ex]{#2}}
\hypersetup{
    colorlinks=true,
    linkcolor=blue,
    filecolor=magenta,      
    urlcolor=blue,
    pdfpagemode=FullScreen,
    }

\DeclareMathOperator{\Term}{\mathcal{T}}
\DeclareMathOperator{\FlatTerm}{\mathcal{FT}}

\DeclareMathOperator{\Cons}{\mathcal{C}}
\DeclareMathOperator{\Kan}{\mathcal{G}}
\DeclareMathOperator{\Fresh}{\mathbf{Fresh}}

\DeclareMathOperator{\Def}{\mathcal{D}}
\DeclareMathOperator{\Base}{\mathbf{Base}}
\DeclareMathOperator{\Conj}{\mathbf{Conj}}

\DeclareMathOperator{\In}{\mathbf{In}}
\DeclareMathOperator{\Out}{\mathbf{Out}}

\newcommand{\KanN}{\mathcal{K}^{N}}

%% file: abstract.tex
\begin{abstract}
  Relational programming enables program synthesis through a verifier-to-solver approach. 
  An earlier paper introduced a functional conversion that mitigated some of the inherent performance overhead. 
  However, the conversion was inelegant: it was oblivious to types, demanded determinism annotations, and implicit generator threading. 
  In this paper, we address these issues by providing a typed tagless-final embedding of \mk into \haskell. 
  This improvement significantly reduces boilerplate while preserving, and sometimes enhancing, earlier speedups. 
\end{abstract}

%% file: introduction.tex
\section{Introduction}

The innate ability of a relational program to run in multiple modes reduces the complicated task of \emph{finding} a solution to the much simpler task of \emph{checking} that a candidate satisfies the specification~\cite{lozov2019relational}. 
For example, a program which verifies whether a given sequence of vertices forms a path in a graph can enumerate paths. 
Likewise, a relation that checks whether some variable assignment satisfies a propositional formula can be run in reverse to search for satisfying assignments. 
A notable instance of this verifier-to-solver approach enables program synthesis: by running a relational interpreter backwards, we can generate programs whose evaluation yields a given value. 

One downside to this approach is its high overhead resulting in low performance. 
A relational interpreter should be carefully crafted to synthesize non-trivial programs within reasonable time constraints. 
Moreover, the implementation of \mk itself has to be heavily optimized which undermines its reputation for being easy to embed in any general-purpose language. 

The \mk community has devoted a lot of effort over the years to make program synthesis a reality.  
Among such efforts is the functional conversion~\cite{engel2023translation} whose purpose is to translate a relation with a given direction into a functional program. 
As a result, much of the overhead of relational programming is neutralized, especially if paired with advanced program transformation techniques such as specialization~\cite{verbitskaia2021deduction}. 

While the implementation of the functional conversion described in~\cite{engel2023translation} improves execution time, it exhibits several limitations. 
Firstly, because it relies on an untyped deep embedding of \mk in \haskell, it fails to take advantage of the \haskell type system, which leads to a loss of static guarantees as well as forces the user to painstakingly thread generator parameters. 
Secondly, the absence of automatic determinism analysis leaves some optimization opportunities untapped. 
Finally, the deep embedded representation hinders composition and extension---qualities that are among \mk's greater strengths. 

In this paper, we describe a typed, tagless-final~\cite{carette2009finally} embedding of \mk in \haskell. 
Along with automatic determinism analysis, it refines the earlier functional conversion, enhancing both performance and developer experience.
In addition to this, the embedding opens the door to composable and extensible \mk implementations in statically typed functional languages.

%% file: background.tex
\section{Background}

In this section we introduce the background necessary to understand the contributions of this paper. 

\subsection{\mk}

The \mk family of programming languages is created to be simple to understand, implement, and extend with new features.\footnote{Website of the \mk programming languages family: \url{http://minikanren.org/}} 
One distinguishing characteristic of \mk is its complete search enabled by design. 
Any existing solution to a query will be found eventually~\cite{rozplokhas2020certified} which makes \mk uniquely suitable for declarative setting and reverse execution. 

Most \mk dialects are implemented as a set of combinators for basic operations such as a unification or conjunction and disjunction. 
However, program transformation is complicated for shallow embeddings in languages featuring referential transparency. 
Primarily due to this fact the proof-of-concept functional conversion was implemented for a deeply embedded \mk and manipulated abstract syntax trees explicitly. 
The syntax in Fig.~\ref{fig:minikanren} describes the core language used in this paper, also known as \micro~\cite{friedman2013mukanren}. 
We also assume inverse $\eta$-delay to accompany every relation invocation and not used elsewhere.  

\begin{figure}[h]
\begin{tabular}{llll}
    $\mathcal{C}$ & $=$ & $\{\Cons_{i}^{k_{i}}\} $ & constructors with arities \\
    $\mathcal{R}$ & $=$ & $\{R_{i}^{k_{i}}\} $ & relational symbols with arities \\
    $\Term_{V}$ & $=$ & $V \cup \{\Cons_{i}\left( t_1, \ldots, t_{k_{i}} \right) \mid t_{j}\in \Term_{V}\} $ & terms on set of variables $V$ \\
 $\Kan_{V}$ & $=$ & $\Term_{V} \equiv \Term_{V}$ & unification \\
 & $\mid$ & $\Kan_{V} \land \Kan_{V}$ & conjunction  \\
 & $\mid$ & $\Kan_{V} \lor \Kan_{V}$ & disjunction  \\
 & $\mid$ & $R_{i}^{k_{i}}\left( t_1, \ldots, t_{k} \right), t_{j}\in \Term_{V}$ & relational symbol invocation\\
 & $\mid$ & $\Fresh_{x} \Kan_{V}$ & fresh variable introduction  \\
 $\Def_{V}$ & $=$ & $R_{i}\left( x_1, \ldots, x_{k_{i}} \right) \equiv \Kan_{V}, x_{j}\in V$ & relational symbol definition \\
  $\mathcal{P}$ & $=$ & $\Def_{V}^*$ & miniKanren program
\end{tabular}
\caption{miniKanren syntax}
\label{fig:minikanren}
\end{figure}

The addition relation written in this syntax takes the following shape, where $R_{add}$ is the only relational symbol, $\Cons_{0}$ and $\Cons_{S}$ are constructors, and $x, x', y, z, z'$ are variables:

\begin{equation*}
    \begin{split}
        R_{\text{add}}\left( x, y, z \right) \equiv 
        &\left(  x \equiv \Cons_0  \land y \equiv z \right) \\
        \lor &\left( \Fresh_{x'}\Fresh_{z'} \left(x \equiv \Cons_{S}\left( x' \right)  \land R_{add}\left(x', y, z'\right) \land z \equiv \Cons_{S}\left( z' \right) \right)\right)
    \end{split}
\end{equation*}

\subsection{Normal Form}
\label{background:normal}

Many program transformations become simpler when done over a normalized representation, such as the superhomogenous normal form used in the Mercury programming language~\cite{mercuryexec}. 
It is in essence a disjunctive normal form with unique variables, and with extra care taken not to allow exponential explosion in the number of clauses. 
To avoid it, we create  a new relation whenever a disjunction takes place within a conjunction. 
The process of normalization is described in detail in \cite{engel2023translation}, here we only present the normal form syntax, see Figure \ref{fig:normalizedminikanren}.
Notice that there are also no $\Fresh$ declarations in this form, and each element of $V$ is assumed to refer to the same variable within the invocation. 

\begin{figure}[h]
\begin{tabular}{llll}
$\FlatTerm_{V}$ & $=$ & $V \cup \{\Cons_{i}\left( x_1, \ldots, x_{k_{i}} \right) \mid x_{j}\in V, a \neq b \implies x_{a} \neq x_{b}\}$ & linear flat terms\\
$\Base_{V}$ & $=$ & $V \equiv \FlatTerm_{V}$ & linear flat unification\\
            & $\mid$ & $R_{i}\left( x_1, \ldots, x_{k_{i}} \right), x_{j}\in V, a \neq b \implies x_{b} \neq x_{b}$ & linear flat call\\
$\Conj_{V}$ & $=$ & $\bigwedge\left( g_1, \ldots, g_n \right), g_{i}\in \Base_{V}$ & normal conjunction \\
$\KanN_{V}$ & $=$ & $\bigvee\left( c_1, \ldots, c_{n} \right), c_{i}\in \Conj_{V}$ & normal form
\end{tabular}
\caption{Normalized miniKanren syntax}
\label{fig:normalizedminikanren}
\end{figure}

\subsection{Modes}

A relational program can be executed in multiple directions, also called \emph{modes}. 
Mode systems assign each variable in a relation a mode annotation that denotes how the variable is used in it. 
A logic/functional programming language \merc features an advanced mode system that informs its compiler on possible optimizations. 
The functional conversion of \mk relies on the simplest mode system with only two possible annotations: $\In$ and $\Out$. 
$\In$ denotes a variable whose value is fully known when the operation is executed. 
$\Out$ is reserved for a variable with unknown value which is going to be bound after the operation is executed. 

When all variables of a relation are annotated with modes, we say that the relation has a mode. 
We use the term \emph{direction} to describe a relation call whose arguments are annotated. 
For example, consider the following directions of the addition relation:

\begin{equation*}
    \begin{split}
        R_{\text{add}}\left( x^{\In}, y^{\In}, z^{\Out} \right) \equiv 
        &\left(  x^{\In} \equiv \Cons_0  \land y^{\In} \equiv z^{\Out} \right) \\
\lor \left( \Fresh_{x'}\Fresh_{z'} \left(x^{\In} \equiv \Cons_{S}\left( x'^{\Out} \right)\right.\right.  \land& \left.\left. R_{add}\left(x'^{\In}, y^{\In}, z'^{\Out}\right) \land z^{\Out} \equiv \Cons_{S}\left( z'^{\In} \right) \right)\right)
    \end{split}
\end{equation*}
\begin{equation*}
    \begin{split}
        R_{\text{add}}\left( x^{\Out}, y^{\Out}, z^{\In} \right) \equiv 
        &\left(  x^{\Out} \equiv \Cons_0  \land y^{\Out} \equiv z^{\In} \right) \\
\lor \left( \Fresh_{x'}\Fresh_{z'} \left(x^{\Out} \equiv \Cons_{S}\left( x'^{\In} \right)\right.\right.  \land& \left.\left. R_{add}\left(x'^{\Out}, y^{\Out}, z'^{\In}\right) \land z^{\In} \equiv \Cons_{S}\left( z'^{\Out} \right) \right)\right)
    \end{split}
\end{equation*}
\begin{equation*}
    \begin{split}
        R_{\text{add}}\left( x^{\In}, y^{\Out}, z^{\Out} \right) \equiv 
        &\left(  x^{\In} \equiv \Cons_0  \land y^{\Out} \equiv z^{\Out} \right) \\
\lor \left( \Fresh_{x'}\Fresh_{z'} \left(x^{\In} \equiv \Cons_{S}\left( x'^{\Out} \right)\right.\right.  \land& \left.\left. R_{add}\left(x'^{\In}, y^{\Out}, z'^{\Out}\right) \land z^{\Out} \equiv \Cons_{S}\left( z'^{\In} \right) \right)\right)
    \end{split}
\end{equation*}

In a well-moded relation, data flows from $\In$ to $\Out$, and once a variable's value is known, it is never forgotten.
This means that a variable with the $\In$ mode can never become $\Out$, while $\Out$ becomes $\In$ after the current operation is executed.
This intuition translates well to functional code, where each variable is expected to be bound by some value before use.
The functional conversion employs a mode analysis capable of inferring mode annotations given a direction.
The mode analysis algorithm does not fall in the scope of this paper; thus, the reader is referred to~\cite{engel2023translation} for a detailed description of the approach.
Its details are not essential for understanding the contributions presented here.

\subsection{Functional Conversion}

In this section we briefly summarize the functional conversion described in~\cite{engel2023translation} which the reader can refer to for deeper comprehension. 
The functional conversion mirrors the relational conversion~\cite{lozov2018typed} aimed at generating relations from functions thus coming full circle in the verifier-to-solver approach.

Given a \mk relation with a concrete direction, the goal of a functional conversion is to construct a function which yields the same answers as the relation would. 
Because the search in \mk is complete, disjunctions and conjunctions can be reordered, which will hopefully result in faster execution. 
Mode analysis guides this reordering as well as classifies unifications into equality checks, assignments, pattern matches and generations. 
Consider the following functional counterpart of $R_{\text{add}}\left( x^{\In}, y^{\Out}, z^{\Out} \right)$ which computes such pairs $(y, z)$ that $x + y = z$, where $x$ is known. 

\includelines{code/addoIOO.hs}{0}{18}

The function \lstinline{addoIOO} produces infinitely many pairs of Peano numbers; thus the output of the function should be a stream of \lstinline{(Term, Term)}. 
The \lstinline{Stream} is not always the most efficient data structure to represent the output, especially when the computation is deterministic. 
Because of this, the converter generates functions parameterized by an arbitrary \lstinline{Monad m} that can be specified at the call site. 

Let us now illustrate the conversion process by this example. 
Once mode analysis is finished, disjunctions are translated into \lstinline{msum} of several monadic computations. 
Each such computation is produced from a conjunction of either calls or unifications. 
Note that moded unifications can become an equality test, a pattern matching, an assignment, or use term generation. 
Generation is required whenever both sides of a unification are annotated $\Out$, producing a stream of all possible values. 

A drawback of the previous implementation is the need to supply generators as explicit function arguments. 
Moreover, when several variables are generated, a separate generator must be provided for each one—even if they share the same type and could be produced in exactly the same way. 
This requirement makes the resulting code inelegant, as well as prone to user errors when they later have to supply the generators manually.

%% file: embedding.tex
\section{Typed Shallow Embedding}
\label{contrib:tagless}

The functional conversion implemented in~\cite{engel2023translation} operates on an untyped version of \mk even though the target languages are \haskell and \ocaml. 
These languages feature expressive static type systems, and it would be a shame to forgo their benefits. 
Besides explicit passing of generators this approach also erases all information about the type of specific variables, and combine all constructors in one type called \lstinline{Term}. 
Figure \ref{fig:generatedterm} shows an example of such type, which integrates natural numbers, lists, and binary trees.
This makes values such as \lstinline{S (Cons (Leaf) (S Nil))} possible, even though they have no meaningful interpretation.  

\begin{figure}[h]
    \centering
    \includelines{code/BalanceoOI.hs}{2}{8} 
    \caption{Generated \emph{Term} type}
    \label{fig:generatedterm}
\end{figure}

To overcome these issues, we need a typed \mk embedding capable of linking each logical value to its ground counterpart. 
Existing candidates fall short.
\haskell embedding typedKanren~\cite{kudasov2024typedkanren} supplies the required logical‑type machinery but only for execution, not for inspecting or analysing relations. 
Simultaneously, \ocanren offers a typed \ocaml embedding yet targets a language less convenient for our conversion pipeline. 
We therefore develop a new, tagless‑final~\cite{carette2009finally} embedding in Haskell that lets programmers write type‑safe and composable relations, provides the correspondence between logic and ground types, and remains open to future extensions. 

The embedding consists of two interfaces expressed as \haskell typeclasses:  

\begin{itemize}
    \item \lstinline{LogicType} that builds correspondence between the logical and the underlying type.
    \item \lstinline{Kanren} that describes a way to interpret a \mk relation. 
\end{itemize}    

These typeclasses are then instantiated to provide concrete logic types and interpreters of relations. 

\subsection{LogicType}
\label{tagless:logic}

\mk interpreters do not operate on concrete \haskell types; instead they expect logical types whose structures mirror the original but contain placeholders for variables, also called \emph{holes}.
The \lstinline{LogicType} typeclass with associated type \lstinline{WithLogic} enforce the shape of a logical type and establish a correspondence between ground and logical types. 
To support multiple interpreters, logic types require variables to represent holes and have to be polymorphic over them, while a variable itself should be polymorphic over the type of its content.
As such, we get the following definition of \lstinline{LogicType} and \lstinline{WithLogic}. 
\includelines{code/Logic.hs}{12}{13}

We also introduce two helper types: \lstinline{Logic}, which represents a logical hole at top-level, and \lstinline{Var}---a variable that contains a logical type.

\includelines{code/Logic.hs}{9}{10}

To illustrate the interface, we will examine it step by step and implement it for a simple case of Peano numbers.
Given the usual implementation of a natural number \lstinline{Nat}, we can define the logic type in the following way. 
\includelines{code/LogicNat.hs}{0}{3}

To map the underlying type to its logical counterpart, \lstinline{LogicType} features functions \lstinline{project} and \lstinline{reify}.

\includelines{code/Logic.hs}{14}{15}

The function \lstinline{project} maps the value of an underlying type to a fully-ground logical value of a logical type, while \lstinline{reify} converts a logical value to its underlying value and returns \lstinline{Nothing} if the input is not fully-ground.
These functions should obey the following laws:

\includelines{code/Logic.hs}{28}{29}

A simple instance for the Peano natural numbers might look like this:
\includelines{code/LogicNat.hs}{15}{20}

To be able to inspect logical values, we provide the following representations of term-style construction. 
It mirrors the typical structure of \mk values: a constructor with a fixed number of arguments of a logical type. 

\includelines{code/Logic.hs}{1}{8}

\lstinline{Field} is a type-erasing GADT, which ensures arguments may have different types. 
\lstinline{Constructor} represents a particular constructor of a term, with an explicit name and a way to create a value.

\lstinline{LogicType} introduces the function \lstinline{quote} which obeys the following law. 
\includelines{code/Logic.hs}{24}{25}

Implementing \lstinline{quote} can be done with helper functions \lstinline{quote0}, \lstinline{quote1}, and others. 
For type-preserving translator, it is important for the names passed to the constructors in \lstinline{quote} to correspond to the constructor names of the underlying type, however it is not in general a requirement of the system.
\includelines{code/LogicNat.hs}{5}{6}

Finally, it should be possible to unify logical types; thus \lstinline{LogicType} contains the function \lstinline{unifyVal}. 
\includelines{code/Logic.hs}{16}{18}

In addition to values to be unified, this function also accepts a \emph{unification provider}.
Unification provider is supplied by the interpreter (see \ref{tagless:interpreters}) and conveys the meaning of the unification in  interpreter's domain.


This function can be implemented for Peano numbers in the following way. 
\includelines{code/LogicNat.hs}{8}{10}


The next function of \lstinline{LogicType} are \lstinline{derefVal} that is similar to \lstinline{reify}, but works on values that can contain variables.
To achieve this, it requires an environment that provides values for such variables through the first argument. 

\includelines{code/Logic.hs}{19}{21} 

Its implementation for natural numbers is as follows. 
\includelines{code/LogicNat.hs}{12}{13}

The goal of the last function, \lstinline{generate}, is to systematically enumerate all possible underlying values. 
It replaces the user-provided generators required in the untyped version \ref{contrib:translator}.

\includelines{code/Logic.hs}{22}{22}

For natural numbers, the generator is rather simple:

\includelines{code/LogicNat.hs}{22}{22}

Functions \lstinline{unifyVal}, \lstinline{derefVal}, and \lstinline{generate} are redundant in the typeclass and can be implemented only by using \lstinline{quote}.
Such an implementation is provided, however its performance may be insufficient for particular interpreters. 
Because of that, the functions are included in the interface and can have custom implementations.

\subsection{Kanren}
\label{tagless:kanren}

The tagless-final approach represent a domain-specific language by an interface of polymorphic combinators. 
Their behavior is determined by interpreters---instances of the typeclass---rather than a fixed syntax tree resulting in extensible, type-safe embeddings. 

In our implementation such interface is called \lstinline{Kanren}.
A type that has a \lstinline{Kanren} instance denotes \emph{relations that return a value}.
Therefore, \lstinline{Kanren rel} means that \lstinline{rel Int} is a relation that produces a value of type \lstinline{Int}. 

Conjunctions are often expressed with a monadic bind, but bind enforces a fixed evaluation order---preventing the reordering of conjuncts that our functional conversion needs.
Instead, we require the \lstinline{Applicative} constraint which imposes no order. 
With this constraint it is impossible to use the value of an earlier conjunct in the following code; thus, for pure \mk relations, the type of the return value is fixed to unit \lstinline{()}. 
The purpose of this parameter is described in \ref{tagless:eval}.

Disjunction, by contrast, naturally fits the \lstinline{Alternative} interface: \lstinline{empty} represents failure, and \lstinline{<|>} models branching without ordering constraints. 
Thus, we simply require \lstinline{Alternative}, which matches the semantics of \mk choice exactly.

Finally, each \lstinline{Kanren} instance has to specify a \lstinline{KVar} type --- a variable associated with this relation. 
It does not need to contain a value, but it has to be functorial over the type it is designated to contain.
Thus, we have the following definition with basic \mk operations:

\includelines{code/Kanren.hs}{6}{9}

In our embedding we need two ways to introduce variables. 
The first one, which corresponds to $\Fresh_{v}$, introduces an empty variable. 
The other denotes an argument of a relation that will, upon introduction, be bound to arguments passed to the call. 
It is equivalent to inserting a unification after $\Fresh_{v}$, but is immediately visible whenever transformations of the relation are done.

\includelines{code/Kanren.hs}{1}{3}

The function \lstinline{fresh} therefore receives a description of the variable to create and a continuation that builds the rest of the relation using that variable, and returns the fully assembled relation: 

\includelines{code/Kanren.hs}{10}{10}

A handful of helpers and the \lstinline{ApplicativeDo} \haskell extension allows writing relations concisely: 
\includecode{code/AddoTF.hs}

\subsection{KanrenEval}
\label{tagless:eval}

Although the \lstinline{Kanren} typeclass makes it possible to define and interpret relations, it offers no mechanism to extract the result of relation execution. 
This may seem counterintuitive, but some interpreters, such as the normalizing interpreter, do not operate with the output, instead transforming the relation itself. 

Nevertheless, it is an important property for evaluators which is why we extend \lstinline{Kanren} with the \lstinline{KanrenEval} typeclass. 
\lstinline{KanrenEval} allows dereferencing variables which makes their value accessible as relation's return value.
An interpreter can then define how to extract the resulting value (see \ref{tagless:interpreters}).

\includelines{code/Kanren.hs}{12}{13}

Combined with \lstinline{derefVal} function from \lstinline{LogicType} (\ref{tagless:logic}), the complete evaluation function is straightforward: 

\includelines{code/Kanren.hs}{15}{17}

\subsection{Automatic Derivation of LogicType Instances}
\label{tagless:types}

\lstinline{LogicType} instances are highly constrained because of the requirement that the logic type mirrors \mk terms and remains interoperable with the underlying type. 
Only algebraic data types (and GADTs) qualify, therefore exactly one valid implementation of \lstinline{LogicType} is admitted per underlying type---a fact we exploit for automatic derivation. 

\lstinline{TemplateHaskell} is capable of generating this unique instance for a given algebraic data type. 
For performance reasons (see Section \ref{task:evaluation}) the implementations of functions in this typeclass must be annotated \lstinline{INLINABLE}. 
This way, the compiler statically resolves which function is to be called for any particular type.

\subsection{Examples of Interpreters}
\label{tagless:interpreters}

Any instance of \lstinline{Kanren} is called an interpreter, whether it evaluates a relation, transforms it, or does something completely different. 
In this subsection we will consider two examples of such interpreters. 

For instance, consider the standard \mk interpreter over a stream of substitutions. 
The result of its execution, or representation, is \lstinline{SubstKanren}. 
We add instances of \lstinline{Kanren} and \lstinline{KanrenEval} for the representation and provide the evaluation function \lstinline{runSubstKanren}. 
\includecode{code/SubstKanren.hs}

Note that the substitution is a partial function that relies on \lstinline{unsafeCoerce}, yet it is safe, as long as \lstinline{SVar} constructor is not used by the programmer. 
Two measures enforce this safety: a marker-parameter \lstinline{s} (inspired by the implementation of \lstinline{ST}), and the type tracking as part of a variable type. 
They ensure that the substitution always produces a value, and \lstinline{unsafeCoerce} is applied only between identical types.

Now let us consider a normalizing interpreter which does not evaluate a relation, instead transforming it. 
It might be encoded in the following way: 
\includecode{code/NormKanren.hs}

We omit the instances of \lstinline{Applicative} and \lstinline{Alternative} which contain the normalization logic, because they are not important to the overall structure. 

\subsection{Extensibility Example}

The major advantage of the tagless-final encoding is that additional structure can be introduced and composed directly in user code without any changes to the framework itself.
Consider normalization from the previous subsection that can transform any relation without changing its observable type.
\includelines{code/NormDemonstration.hs}{1}{9}

However, here normalization is totally transparent. 
Even though the structure of the relation is more constrained, it is not exposed and cannot be used in further transformations, because the only interface available is \lstinline{Kanren}.
We can combat this issue by defining a new typeclass, \lstinline{NormalizedKanren}, to expose the normalized shape and make it interoperable with the \lstinline{Kanren} typeclass.

\includecode{code/NormKanrenTypeclass.hs}

With this setup, any relation can be processed by a normalizing interpreter before being passed on to a conversion which will take full advantage of the normalized structure. 
Conversely, with an instance of \lstinline{NormalizedKanrenT}, any normalized relation can be restored to the \lstinline{Kanren} typeclass with the \lstinline{normalize} function, erasing the structure.
As a result, we demonstrated extensibility by providing two interoperable relation representations with no need to modify \lstinline{Kanren}, or any other core feature.

%% file: conversion.tex
\section{Type-Preserving Functional Conversion}
\label{contrib:translator}

In this section, we present the updated functional-conversion pipeline that takes advantage of the typed \mk embedding. 
Rather than rebuilding the converter from scratch, we implemented an interpreter that lifts a shallowly embedded, typed relation into the deep, untyped representation used in the earlier work~\cite{engel2023translation}. 
This choice lets us reuse the existing mechanisms of normalization, mode and determinism analyses, while still preserving limited type safety. 

Even though the internal representation remains untyped, the fact that the translated relation originates from a typed embedded means that its underlying \haskell types as well as generators are available. 
This eliminates some inelegant aspects of the conversion process, namely synthesizing a monolithic \emph{Term} type of all constructors, and threading explicit generators. 
Because of our assumption that the \emph{quote} function (see \ref{tagless:logic}) uses the actual constructor names of the underlying type, the converter can invoke them directly and call the type's \emph{generate} whenever enumeration is required.

As an illustration, consider the relation $R_{\text{balance}}(x, y)$ that holds when the binary tree $y$ is the balanced version of $x$. 

\includecode{code/Balanceo.hs}

We translate this relation in mode  $R_{\text{balance}}\left( x^{\Out}, y^{\Out} \right)$, thus enumerating all trees alongside their balanced versions. 
The earlier functional conversion produces the following result. 
In addition to producing the monolithic \emph{Term} type, it demonstrates the generator threading: \emph{balancedoO\_x2} is not used in \emph{balanceoOO}, instead it is passed to \emph{balancedoO}.
Finally, it misses an optimization opportunity in line 14, where the \emph{Stream} monad is used instead of the semi-deterministic \emph{Maybe}. 

\includecode{code/BalanceoOI.hs}

By contrast, the type-preserving translation avoids creating the \emph{Term} type and relies on the underlying types' constructors and the functions \emph{generate}.
An automatic determinism analysis allows us to use the \emph{Maybe} monad whenever we can detect semi-deterministic computations. 
These computations are executed faster in \emph{Maybe} and then lifted back to the \emph{Stream} monad by \emph{liftMaybe}. 

\includecode{code/BalanceoTI.hs}

%% file: experiments.tex
\section{Experiments}
\label{task:evaluation}

To gauge the impact of our changes, we tested our functional conversion (New) against the untyped one (Old) on several relations. 
The new pipeline matches the old in every case and improves upon it in presence of semi-deterministic relations. 
We consider three relations: addition, generating permutations as a reverse of sorting, and typechecking used to enumerate terms of a given type. 
These relations showcase different aspects of performance gains.

\begin{figure}[ht]
\makebox[\textwidth][c]{\includegraphics[width=0.9\textwidth]{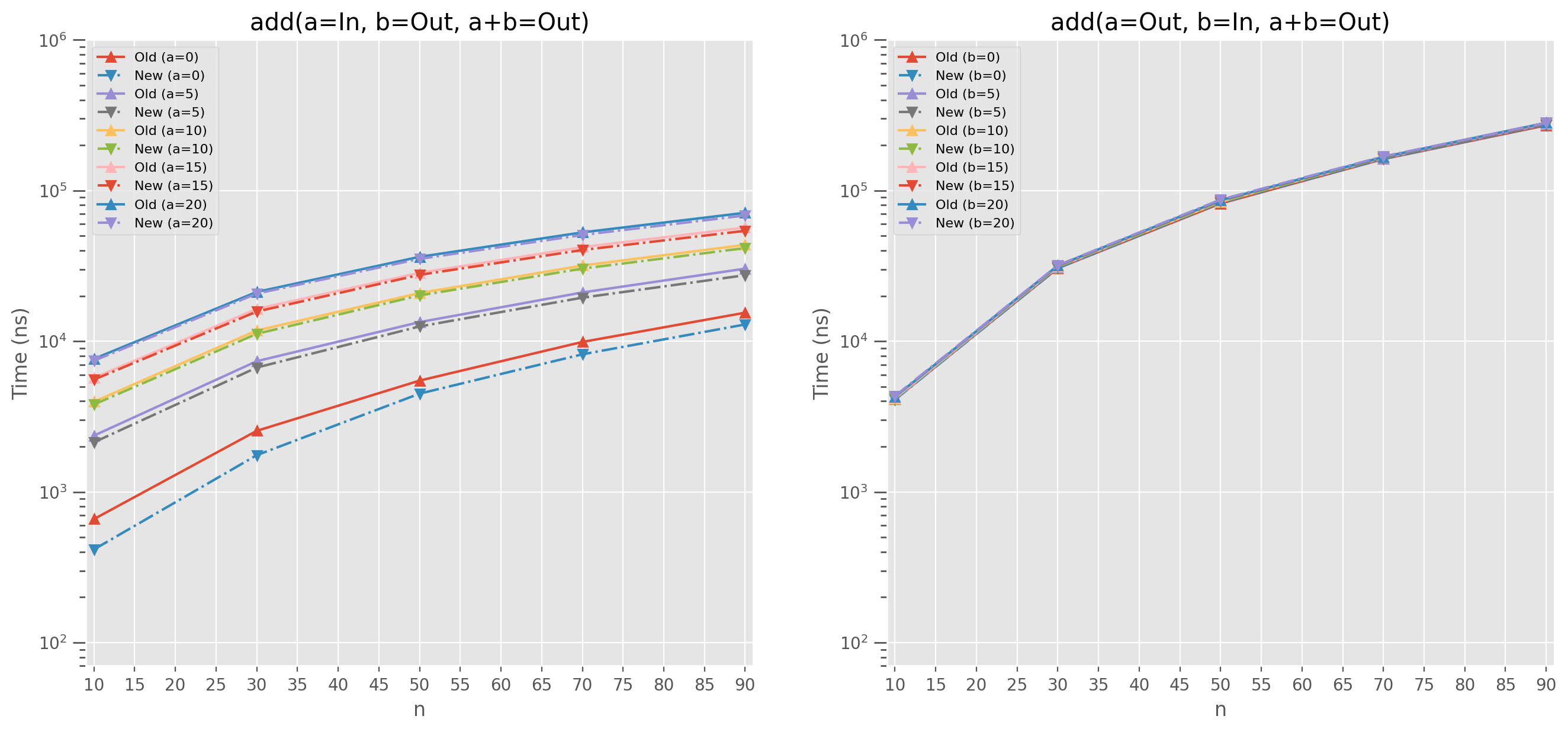}}
\caption{Execution time (in nanoseconds) of $add$ relation in two nondeterministic directions, requesting first $n$ (x-axis) results. Log plot.}
\label{fig:addNondet}
\end{figure}

\begin{figure}[h!]
\makebox[\textwidth][c]{\includegraphics[width=0.9\textwidth]{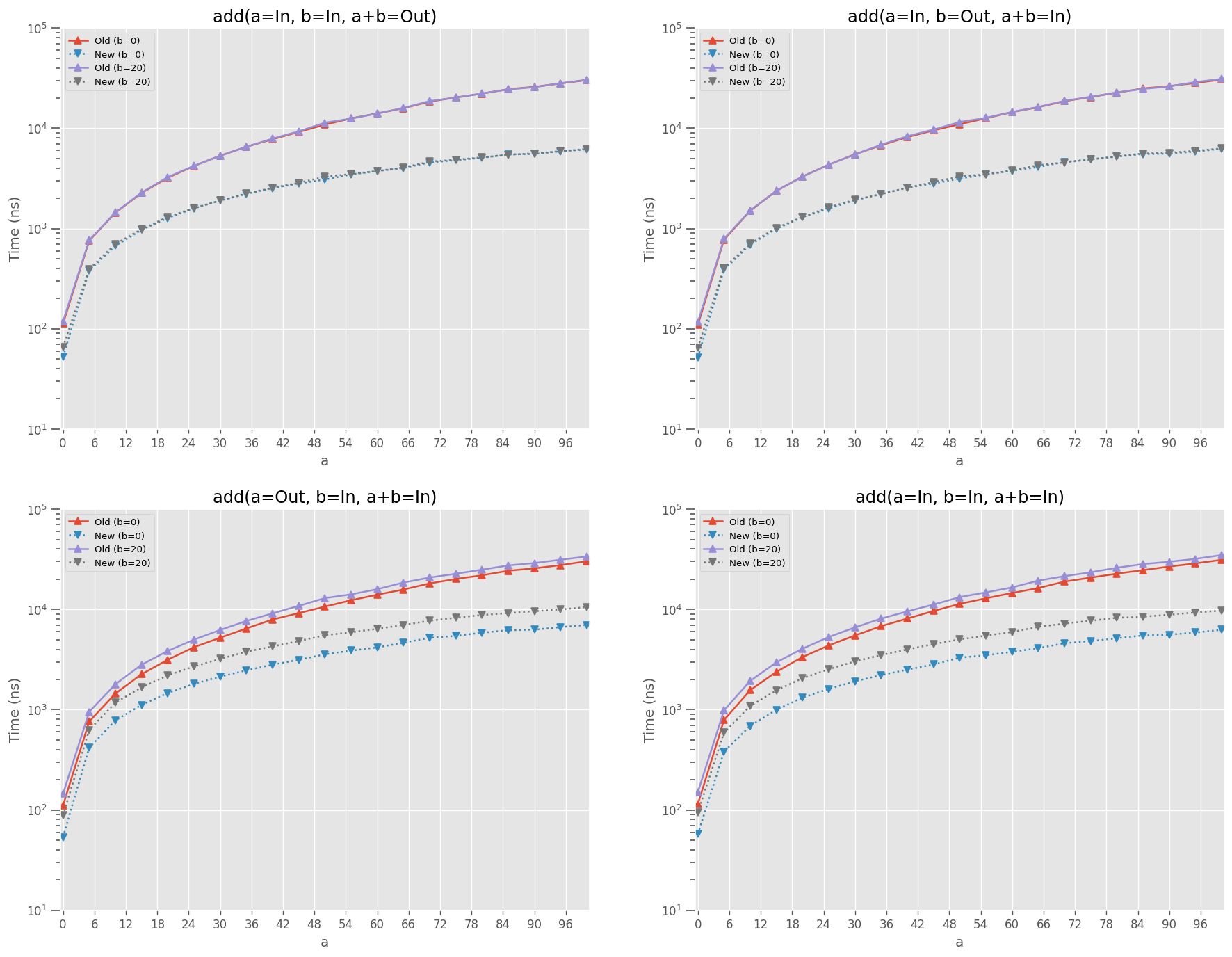}}
\caption{Execution time (in nanoseconds) of $add$ relation in four deterministic directions, with parameter $a$ on x-axis. Log plot.}
\label{fig:addDet}
\end{figure}

Figure \ref{fig:addNondet} presents nondeterministic directions of addition, when exactly one variable is known. 
In the direction $R_{\text{add}}\left( \In, \Out, \Out \right)$ the relation uses one generator, and shows slight speed-up. 
The $R_{\text{add}}\left( \Out, \In, \Out \right)$ direction does not feature any optimization opportunities for our improved converter since it is nondeterministic, uses a single type, and avoids using generators. 
As a result, its performance matches the old implementation exactly.

Figure \ref{fig:addDet} covers semi-deterministic directions of addition.
A clear performance improvement can be observed across all functions,
reaffirming the importance of tracking determinism.

\begin{figure}[ht]
\makebox[\textwidth][c]{\includegraphics[width=0.9\textwidth]{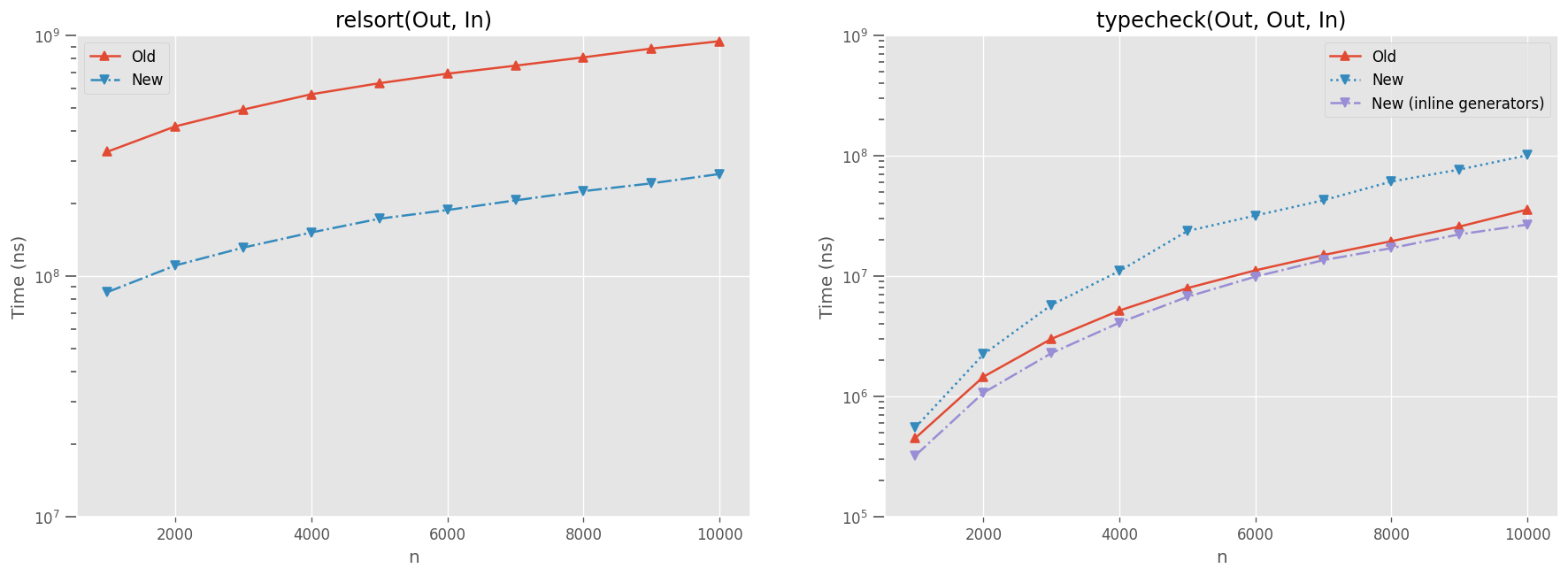}}
\caption{Execution time (in nanoseconds) of generating permutations from $12$-element list with the $sort$ relation and enumerating terms and contexts that infer to a given type by the $typecheck$ relation, requesting first $n$ (x-axis) results. Log plot.}
\label{fig:sorttypecheck}
\end{figure}

The left panel of Figure \ref{fig:sorttypecheck} showcases generating permutations of a sorted $12$-element list by the \emph{sort} relation. 
We observe performance improvement which arises entirely from semi-determinism of comparison relations.

The right panel of Figure \ref{fig:sorttypecheck} utilizes the $typecheck$ relation to enumerate terms of a simple expression language that type check to the \emph{Integer} type. 
This example demonstrates the importance of using \lstinline{INLINEABLE} annotation for generators. 
When they are correctly inlined, we observe a reduced execution time, while naive implementation only worsens performance. 

Overall, the new converter never slows down the relations. 
Moreover, in some cases, determinism analysis yields qualitative improvements by identifying computations which can safely be pruned once a single answer has been produced. 

%% file: conclusion.tex
\section{Conclusion}

We presented a typed tagless-final embedding of \mk in \haskell and a functional conversion based on it. 
It addresses several inefficiencies and inelegant design decisions of the previous implementation, including monolithic \emph{Term} types, generation threading and explicit determinism annotations. 
Moreover, the embedding promotes extensibility, one of the important aspects of the \mk language family.